\documentclass[pdftex,twocolumn,epjc3]{svjour3}          

\usepackage{array, tabularx, caption, boldline}
\usepackage{graphicx}
\usepackage{makecell}
\usepackage{cellspace}
\usepackage[utf8]{inputenc}
\usepackage{lmodern,textcomp}

\usepackage{float}
\usepackage{cite}
\usepackage{ifthen, mciteplus} 

\usepackage{url}
\newboolean{pdflatex}
\setboolean{pdflatex}{true} 

\newboolean{articletitles}
\setboolean{articletitles}{true} 
\newboolean{uprightparticles}
\setboolean{uprightparticles}{false} 
\newboolean{inbibliography}
\setboolean{inbibliography}{true} 

\usepackage{lineno}  
\usepackage{xspace} 
\usepackage{caption} 

\RequirePackage[T1]{fontenc}

\smartqed  

\RequirePackage{graphicx}
\RequirePackage{mathptmx}      
\RequirePackage{flushend}
\RequirePackage[numbers,sort&compress]{natbib}
\RequirePackage[colorlinks,citecolor=blue,urlcolor=blue,linkcolor=blue]{hyperref}

\journalname{Eur. Phys. J. C}

\usepackage{xspace} 
\usepackage{upgreek}
\usepackage{amsmath}

\def\DSL   {\ensuremath{X_{b}\to (X_{c} \to K^{+} \ell^{-} \nu_{\bar{\ell}} X) \ell^{+} \nu_{\ell} X}\xspace}
\def\bKll   {\ensuremath{X_{b}\to K^{+} \ell^{+}\ell^{-} X}\xspace}
\def\incl   {\ensuremath{\Bbar\to X_{s} \ell^{+}\ell^{-} }\xspace}
\def\bsll   {\ensuremath{b\to s \ell^{+}\ell^{-}}\xspace}

\def\MagUp {\mbox{\em Mag\kern -0.05em Up}\xspace}

\ifthenelse{\boolean{uprightparticles}}%
{

 \def\Pmu         {\ensuremath{\upmu}\xspace}

 \def\Ppi         {\ensuremath{\uppi}\xspace}

 \def\Ppsi        {\ensuremath{\uppsi}\xspace}

 \def\PDelta      {\ensuremath{\Delta}\xspace}                 
 \def\PXi         {\ensuremath{\Xi}\xspace}                 
 \def\PLambda     {\ensuremath{\Lambda}\xspace}                 
 \def\PSigma      {\ensuremath{\Sigma}\xspace}                 
 \def\POmega      {\ensuremath{\Omega}\xspace}                 
 \def\PUpsilon    {\ensuremath{\Upsilon}\xspace}

 \def\PB      {\ensuremath{\mathrm{B}}\xspace}                 
                  
 \def\PD      {\ensuremath{\mathrm{D}}\xspace}

 \def\PJ      {\ensuremath{\mathrm{J}}\xspace}                 
 \def\PK      {\ensuremath{\mathrm{K}}\xspace}

 \def\Pb      {\ensuremath{\mathrm{b}}\xspace}

 \def\Pe      {\ensuremath{\mathrm{e}}\xspace}

 \def\Pi      {\ensuremath{\mathrm{i}}\xspace}

 \def\Ps      {\ensuremath{\mathrm{s}}\xspace}

 \def\thebaroffset{0.0em}
}
{

 \def\Pmu         {\ensuremath{\mu}\xspace}

 \def\Ppi         {\ensuremath{\pi}\xspace}

 \def\Ppsi        {\ensuremath{\psi}\xspace}                 
                  
 \mathchardef\PDelta="7101
 \mathchardef\PXi="7104
 \mathchardef\PLambda="7103
 \mathchardef\PSigma="7106
 \mathchardef\POmega="710A
 \mathchardef\PUpsilon="7107
                  
 \def\PB      {\ensuremath{B}\xspace}                 
                  
 \def\PD      {\ensuremath{D}\xspace}

 \def\PJ      {\ensuremath{J}\xspace}                 
 \def\PK      {\ensuremath{K}\xspace}

 \def\Pb      {\ensuremath{b}\xspace}

 \def\Pe      {\ensuremath{e}\xspace}

 \def\Pi      {\ensuremath{i}\xspace}

 \def\Ps      {\ensuremath{s}\xspace}

 \def\thebaroffset{0.18em}
}
\newcommand{\offsetoverline}[2][\thebaroffset]{\kern #1\overline{\kern -#1 #2}}%

\makeatletter
\ifcase \@ptsize \relax
  \newcommand{\miniscule}{\@setfontsize\miniscule{4}{5}}
\or
  \newcommand{\miniscule}{\@setfontsize\miniscule{5}{6}}
\or
  \newcommand{\miniscule}{\@setfontsize\miniscule{5}{6}}
\fi
\makeatother

\DeclareRobustCommand{\optbar}[1]{\shortstack{{\miniscule (\rule[.5ex]{1.25em}{.18mm})}
  \\ [-.7ex] $#1$}}

\def\epem       {{\ensuremath{\Pe^+\Pe^-}}\xspace}

\def\mumu       {{\ensuremath{\Pmu^+\Pmu^-}}\xspace}

\def\squark    {{\ensuremath{\Ps}}\xspace}

\def\bquark    {{\ensuremath{\Pb}}\xspace}

\def\pion   {{\ensuremath{\Ppi}}\xspace}

\def\pip    {{\ensuremath{\pion^+}}\xspace}
\def\pim    {{\ensuremath{\pion^-}}\xspace}

\def\kaon    {{\ensuremath{\PK}}\xspace}

\def\KorKbar {\kern \thebaroffset\optbar{\kern -\thebaroffset \PK}{}\xspace}

\def\Kp      {{\ensuremath{\kaon^+}}\xspace}
\def\Km      {{\ensuremath{\kaon^-}}\xspace}

\def\Kstarz  {{\ensuremath{\kaon^{*0}}}\xspace}

\def\DorDbar {\kern \thebaroffset\optbar{\kern -\thebaroffset \PD}\xspace}

\def\B       {{\ensuremath{\PB}}\xspace}
\def\Bbar    {{\ensuremath{\offsetoverline{\PB}}}\xspace}

\def\BorBbar {\kern \thebaroffset\optbar{\kern -\thebaroffset \PB}\xspace}
\def\Bz      {{\ensuremath{\B^0}}\xspace}

\def\Bd      {{\ensuremath{\B^0}}\xspace}

\def\BdorBdbar {\kern \thebaroffset\optbar{\kern -\thebaroffset \Bd}\xspace}
\def\Bu      {{\ensuremath{\B^+}}\xspace}

\def\Bp      {{\ensuremath{\Bu}}\xspace}

\def\Bs      {{\ensuremath{\B^0_\squark}}\xspace}

\def\BsorBsbar {\kern \thebaroffset\optbar{\kern -\thebaroffset \Bs}\xspace}

\def\jpsi     {{\ensuremath{{\PJ\mskip -3mu/\mskip -2mu\Ppsi\mskip 2mu}}}\xspace}

\def\Y#1S{\ensuremath{\PUpsilon{(#1S)}}\xspace}

\def\Lz          {{\ensuremath{\PLambda}}\xspace}

\def\LorLbar     {\kern \thebaroffset\optbar{\kern -\thebaroffset \PLambda}\xspace}

\def\Lb           {{\ensuremath{\Lz^0_\bquark}}\xspace}

\newcommand{\decay}[2]{\ensuremath{#1\!\to #2}\xspace} 

\def\to                 {\ensuremath{\rightarrow}\xspace}

\def\CP                {{\ensuremath{C\!P}}\xspace}

\def\bsll     {\decay{\bquark}{\squark \ell^+ \ell^-}}

\def\AT#1     {\ensuremath{A_{\mathrm{T}}^{#1}}\xspace}           

\def\C#1      {\ensuremath{\mathcal{C}_{#1}}\xspace}                       
\def\Cp#1     {\ensuremath{\mathcal{C}_{#1}^{'}}\xspace}                    
\def\Ceff#1   {\ensuremath{\mathcal{C}_{#1}^{\mathrm{(eff)}}}\xspace}        
\def\Cpeff#1  {\ensuremath{\mathcal{C}_{#1}^{'\mathrm{(eff)}}}\xspace}       
\def\Ope#1    {\ensuremath{\mathcal{O}_{#1}}\xspace}                       
\def\Opep#1   {\ensuremath{\mathcal{O}_{#1}^{'}}\xspace}                    

\newcommand{\aunit}[1]{\ensuremath{\text{\,#1}}}       

\newcommand{\tev}{\aunit{Te\kern -0.1em V}\xspace}
\newcommand{\gev}{\aunit{Ge\kern -0.1em V}\xspace}
\newcommand{\mev}{\aunit{Me\kern -0.1em V}\xspace}
\newcommand{\kev}{\aunit{ke\kern -0.1em V}\xspace}
\newcommand{\ev}{\aunit{e\kern -0.1em V}\xspace}
\newcommand{\mevc}{\ensuremath{\aunit{Me\kern -0.1em V\!/}c}\xspace}
\newcommand{\gevc}{\ensuremath{\aunit{Ge\kern -0.1em V\!/}c}\xspace}
\newcommand{\mevcc}{\ensuremath{\aunit{Me\kern -0.1em V\!/}c^2}\xspace}
\newcommand{\gevcc}{\ensuremath{\aunit{Ge\kern -0.1em V\!/}c^2}\xspace}

\def\fb   {\ensuremath{\aunit{fb}}\xspace}
\def\invfb   {\ensuremath{\fb^{-1}}\xspace}

\def\gsim{{~\raise.15em\hbox{$>$}\kern-.85em
          \lower.35em\hbox{$\sim$}~}\xspace}
\def\lsim{{~\raise.15em\hbox{$<$}\kern-.85em
          \lower.35em\hbox{$\sim$}~}\xspace}

\def\tell1  {TELL1\xspace}
\def\ukl1   {UKL1\xspace}

\newcommand{\ie}{\mbox{\itshape i.e.}\xspace}

\begin{document}

\title{Isospin extrapolation as a method to study inclusive \incl decays}

\author{Yasmine Amhis\thanksref{e1,addr1}
        \and
        Patrick Owen\thanksref{e2,addr2} 
}

\thankstext{e1}{e-mail: yasmine.amhis@ijclab.in2p3.fr}
\thankstext{e2}{e-mail: powen@physik.uzh.ch}

\institute{Universit\'e Paris-Saclay, CNRS/IN2P3, IJCLab, 91405 Orsay, France\label{addr1}
          \and
          Physik-Institut, Universit\"at Z\"urich, CH-8057 Z\"urich, Switzerland\label{addr2}
}

\date{Received: June 23, 2021 / Accepted: date}

\maketitle

\begin{abstract}
A novel approach to reconstruct inclusive \incl decays is presented. The method relies on isopsin symmetry to extrapolate the semi-inclusive signature \bKll to the fully inclusive rate in \Bp and \Bz decays.  We investigate the possibility to measure branching fractions and other observables such as lepton universality ratios and \CP asymmetries. As a proof of concept, fast simulation is used to compare the \bKll signature with a fully inclusive approach. Several experimental advantages are seen which have the potential to make measurements of inclusive \incl decays tractable at a hadron collider.
\end{abstract}

\section{Introduction}
\label{sec:introduction}
The last few years have seen several deviations in measurements of rare \bsll decays\footnote{The inclusion of charge conjugated processes is assumed throughout.} with respect to their Standard Model (SM) predictions~\cite{Aaij:2013qta,Aaij:2014ora,CMS:2014xfa,Aaij:2015oid,Aaij:2017vbb,Aaboud:2018mst,Sirunyan:2019xdu,Aaij:2019wad,Aaij:2020nrf,Aaij:2020ruw,Aaij:2021vac}. The combined statistical significance of these anomalies is large~\cite{Alguero:2021anc,Altmannshofer:2021qrr,Geng:2021nhg}, but partially depends on theoretical uncertainties which are under debate~\cite{Lyon:2014hpa,Bharucha:2020eup,Gubernari:2020eft,Huber:2019iqf,Jager:2019bgk,Huber:2019iqf,Nakayama:2019eth,Arbey:2018ics,Chrzaszcz:2018yza,Blake:2017fyh,Ahmady:2015fha,Descotes-Genon:2015xqa,Ciuchini:2015qxb}.
Some deviations, such as lepton universality measurements~\cite{Aaij:2019bzx, Aaij:2017vbb, Aaij:2021vac} and the $\Bs\to\mumu$ branching fraction~\cite{Aaboud:2018mst,Sirunyan:2019xdu,Aaij:2017vad} are not limited by theoretical uncertainties. However, the significance of these measurements is not yet above the 5$\sigma$ discovery threshold~\cite{Lancierini:2021sdf,Alguero:2021anc,Altmannshofer:2021qrr,Geng:2021nhg,Hurth:2021nsi,Cornella:2021sby}. The motivation for more measurements which probe the same underlying quark transition but have different theoretical and experimental uncertainties is therefore clear. 

The  experimental precision of \bsll decays is currently dominated by measurements of \emph{exclusive} decays, whereby the strange quark hadronises into a specific final state. Exclusive measurements have the advantage of low backgrounds and a good resolution of the full decay kinematics. However, their interpretation, particularly for branching fractions,  suffers from theoretical uncertainties associated to hadronic form factors~\cite{Straub:2015ica,Bailey:2015dka,Du:2015tda,Horgan:2013pva,Horgan:2015vla} which in some cases saturate the uncertainty between the SM predictions and the experimental measurements~\cite{Aaij:2014pli,Aaij:2016flj,Aaij:2021pkz}. 

This paper describes an approach to measure inclusive \incl decays, whereby the strange quark is allowed to hadronise to any final state. Inclusive \incl measurements have complementary theoretical uncertainties to those of exclusive decays~\cite{Huber:2019iqf,Hurth:2010tk,Huber:2015sra,Huber:2020vup}.  The most precise measurements~\cite{Lees:2013nxa,Iwasaki:2005sy,Sato:2014pjr} originate from the $B$-factories, and are performed using a sum-of-exclusives approach, whereby several exclusive final states are combined and extrapolated to the full inclusive rate using a hadronisation model. In addition to this, a selection requirement is applied to the mass of the strange hadron, $m_{X_{s}}$. This is necessary to reduce background but requires an extrapolation to the full mass range using theoretical models.

The ideal reconstruction method to avoid such extrapolations is to apply a fully inclusive approach and only reconstruct the two leptons~\cite{Kou:2018nap}. Such a technique is difficult experimentally, particularly at a hadron collider due to the presence of large amounts of partially reconstructed backgrounds, which are more easily confused with inclusive signal decay of interest.

The purpose of this paper is to describe a new approach to measure inclusive decays, which is to reconstruct a single charged kaon in addition to the two leptons. The presence of the additional kaon is expected to provide several experimental advantages with respect to a fully inclusive approach. The interpretation the branching fraction of a \bKll signature requires an extrapolation to the fully inclusive rate via isospin rules to include the part with neutral kaons. Such an extrapolation is expected to have different theoretical uncertainties with respect to extrapolating a sum-of-exclusives method and should therefore provide a complementary measurement to the existing ones.

The paper is structured as follows: First we describe the experimental method and the advantages with respect to a fully inclusive approach, then we discuss potential complications with the extrapolation to the fully inclusive rate. Finally, we briefly describe prospects for observables which have theoretically precise SM predictions, which can be interpreted even under the application of stringent kinematic selection. 

\section{Experimental signature}
\label{sec:method}

We propose to combine a charged kaon with two leptons to form an inclusive \bKll signature, where $X_{b}$ is a ground-state $b$-hadron and $X$ represents a number of unreconstructed particles. In order to preserve the inclusive signature, no impact parameter requirements should be placed on the combination of the visible products, nor should any track or vertex isolation requirements be applied. This leaves the analysis open to partially reconstructed backgrounds, whereby additional particles not reconstructed can be easily confused with the inclusive signal.

The most dangerous background arises from double semileptonic decays, where both a $b$ and $c$-hadron decay semileptonically, resulting in a \DSL signature. At least two neutrinos will be missing which can easily be confused with the missing particles from the $X_{s}$ hadron of the signal decay. There are a couple of features which can distinguish this background from the signal. Firstly, the lifetime of the charm hadron will result in a poorer quality vertex compared to signal. This can be exploited without concern of systematic bias towards particular strange meson species. Secondly, there will be a momentum imbalance between the two leptons due to the different kinematics of the charm and beauty hadrons. This will be highly discriminating against the background, with the price of a systematic uncertainty associated to a kinematic selection of the signal. 

Another source of contamination to be treated at a hadron collider would be that from combinatorial background, whereby accidental combinations of particles from different $b$-hadron decays are made. Here there will be various kinematic and geometric features which can be exploited. However, the content of combinatorial background is generally less understood than the double-semileptonic, usually mandating a sideband in data. For fully reconstructed decays, the sideband is in close proximity to the signal due to the narrow invariant mass resolution. For inclusive decays, however, the sideband can be far from the signal region as it implies a region that is kinematically forbidden for signal.

\begin{figure}[tb]
    \centering
    \includegraphics[width=0.45\textwidth]{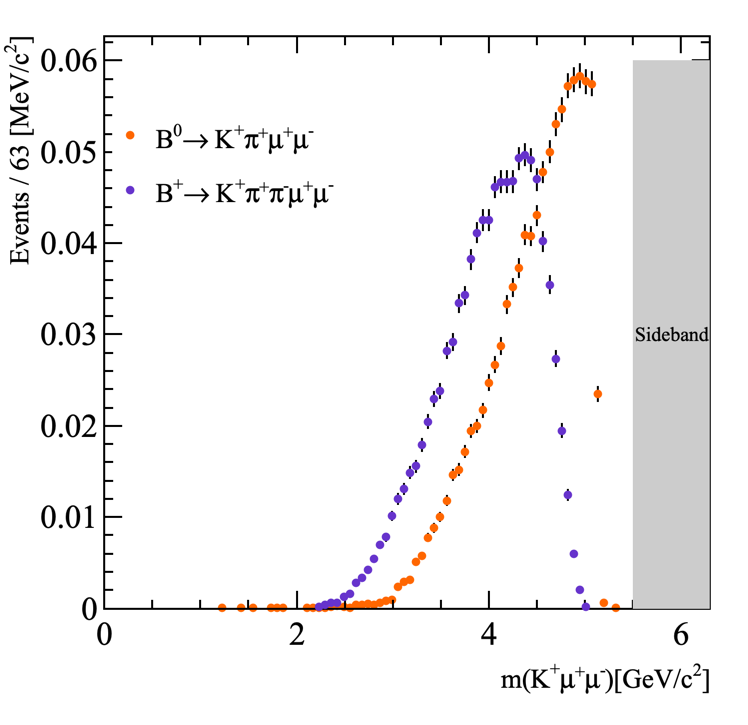}
    \includegraphics[width=0.45\textwidth]{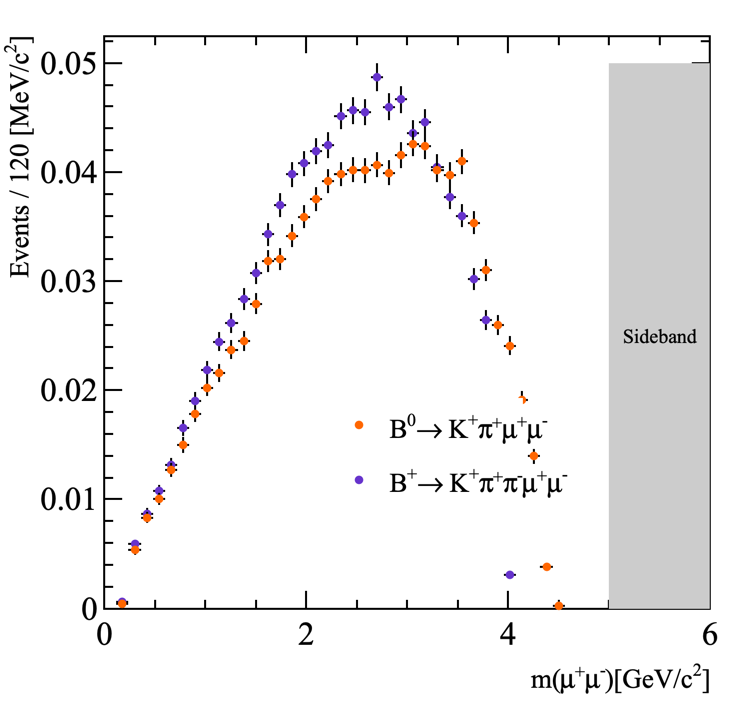}
    \caption{Comparison of the $\Kp\mumu$ invariant mass combination (left) and  di-lepton invariant mass (right) for partially  reconstructed  $\Bz \to \Kp \pim  \mu^{+}\mu^{-}$ and $\Bp\to ( K_{1}(1270)^{+}\to \Kp\pim\pip) \mu^{+}\mu^{-}$ decays. The sideband region is also shown. The greyed out areas correspond to the sideband. These distributions were produced using the fast simulation~\cite{Cowan:2016tnm} described in the text.}
     \label{fig:visiblemass}
\end{figure}

\begin{figure}[tb]
    \centering
    \includegraphics[width=0.45\textwidth]{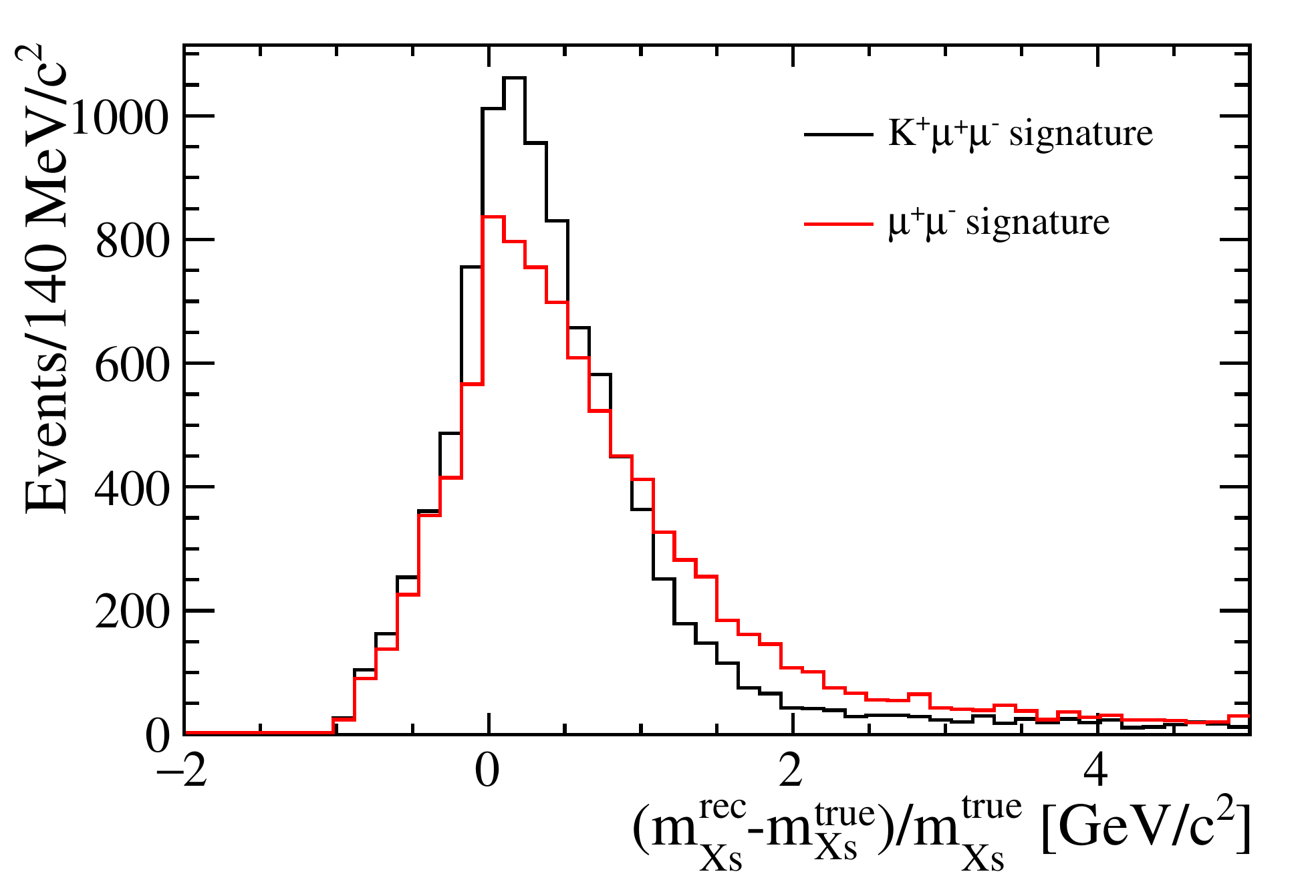}
    \caption{Resolution on the invariant mass of the strange hadron, $m_{X_{s}}$ comparing when only the two leptons are reconstructed and when an additional kaon is reconstructed for $\Bp\to ( K_{1}(1270)^{+}\to \Kp\pim\pip) \mu^{+}\mu^{-}$ decays.}
    \label{fig:mXs}
\end{figure}

 In order to investigate some of these issues, a fast simulation~\cite{Cowan:2016tnm} is performed, where $b$-hadrons are simulated using at 13 TeV inside the pseudo-rapidity range $2<\eta<5$, approximately corresponding to the LHCb geometrical acceptance. All the reconstructed final state particles \ie the kaon and the two leptons are required to have a transverse momentum larger than 300 MeV/$c^{2}$. By construction this fast simulation does not account for all the possible detector effects that will be present, however it is enough for our proof of concept. The decays $\Bz\to (\Kstarz\to\Kp\pim)\mu^{+}\mu^{-}$ and $\Bp\to (K_{1}(1270)^+\to\Kp\pim\pip)\mu^{+}\mu^{-}$ are used as a proxy for a typical partially reconstructed decay, where the pions are not reconstructed. The background is the largest issue to solve in the experimental analysis. This is where the reconstruction of an additional kaon offers a few experimental advantages in this regard:
\begin{itemize}
    \item Reconstructing an additional kaon provides a more well-defined vertex with respect to a di-lepton vertex. This allows to more easily suppress double-semileptonic background due to the finite lifetime of the charm hadron which would result in a poorer vertex quality. It will also improve suppression of combinatorial background. The precise gain would need to be studied with full simulation.
    \item The invariant mass of the kaon and the opposite sign lepton ($m_{K^{+}\mu^{-}}$) pair can be calculated. For the double-semileptonic background, this must be below the mass of the $D$ meson which allows for a good discrimination power.
    \item The visible mass $m_{\Kp\ell^{+}\ell^{-}}$ allows for a much smaller extrapolation from the signal region to the sideband. An example of this is shown in Fig.~\ref{fig:visiblemass}, which compares the distribution for this visible mass with $q^{2}$ for partially reconstructed signal decays. 
    \item The mass of the strange meson, $m_{X_{s}}$ can be calculated using the rest frame approximation~\cite{Aaij:2015yra}. This is another aspect which can be used to discriminate between signal and background. With the addition of the charged kaon, the resolution of this variable improves as can be seen in Fig.~\ref{fig:mXs}.
    \item The signature is self-tagged, meaning that $CP$ asymmetries and the forward-backward asymmetry, such as in Ref.~\cite{Sato:2014pjr}, can be measured without flavour tagging. 
\end{itemize}

These experimental advantages clearly demonstrate the potential of reconstructing a charged kaon in addition to the two leptons. Given that the signal uncertainty is currently the limiting factor in the measurements, this should be improve the overall impact of inclusive decays in the experimental precision of \bsll decays.

\section{Comments on the extrapolation}
\label{sec:extrapolation}

The interpretation of the branching fraction of \bKll decays relies on an extrapolation to the fully inclusive \incl rate in order to compare with theoretical predictions. The question reduces down to the probability that the inclusive \incl decay includes a charged kaon in its decay products. For \Bp  and \Bz decays, it is expected that the inclusive \incl signature will always result in a charged or neutral ground-state kaon. The key to extrapolate to the fully inclusive rate is then the use of isospin symmetry, which is known to hold well for the \Bp  and \Bz decays~\cite{Khodjamirian:2012rm,Lyon:2013gba}. 

At the LHC, around 10\% of ground state $\bquark$-hadrons are \Bs mesons~\cite{Aaij:2019pqz}. Isospin symmetry should also hold well for the inclusive $\Bs\to \bar{X}_{s\bar{s}}\ell^{+}\ell^{-}$ decay to account for neutral kaons. However, there are several mesons with a non-zero $s\bar{s}$ component such as $\eta^{(')}$ mesons. Such decays will not result in ground state kaon and would have to be accounted for either theoretically or by specifically reconstructing the exclusive final states. 

The other significant contribution to the $\bquark$-hadron production at the LHC is that from \Lb baryons, which make up around 20\% of the total $b\bar{b}$ rate. The complication here arises from the $\Lz$ baryon family, where the ground state does not decay into a kaon and the excited states decay into several different final states including \Lz, $\Delta$, $\Sigma$ baryons. An extrapolation to the inclusive $\Lb\to \Lambda^{(*)} \ell\ell$ decay therefore appears difficult.

Compared to the \Bp and \Bz mesons, the \Bs and \Lb decays have a significantly smaller contribution to the \bKll signature due to their smaller production and branching fractions. Our proposal is to therefore treat the contributions from the \Bs and \Lb hadrons as background, and therefore subtract their contribution using dedicated measurements. Such measurements can include the recent exclusive $\Bs\to\phi\mumu$ analysis~\cite{Aaij:2021pkz} and $\Lb\to p\Km\ell^{+}\ell^{-}$ analysis~\cite{Aaij:2019bzx}. Dedicated inclusive measurements such as $\Bs\to\Kp\Km\ell^{+}\ell^{-} X$ and $\Lb\to p \Km \ell^{+}\ell^{-} X$ could also be considered as they will be  helpful to account for residual background contributions. 
In any case, the abundant $X_{b} \to \jpsi X$ modes will be a valuable control channel to check such an extrapolation even if the hadronic dynamics will be different to the rare mode. 

\section{Prospects for theoretically precise observables}
\label{sec:extrapolation}

The main motivation for the isospin extrapolation method is to determine the branching fraction of the inclusive \incl decay. However, there are other observables which can benefit from the large signal yields expected with inclusive decays. Such observables include searches for the lepton flavour violating decay $b\to s e^{\pm} \mu^{\pm}$, the test of lepton universality $R_{KX}=\Gamma(X_{b}\to K\mu^{+}\mu^{-} X)/\Gamma(X_{b}\to K e^{+} e^{-} X)$ and measurements of the \CP asymmetry. Electron reconstruction is challenging at a hadron collider~\cite{Aaij:2019vvl}. However, the experimental control should be similar to those seen in the exclusive analyses. The SM prediction for these observables is expected to be theoretically precise even with a partial reconstruction of the inclusive decay rate.


Inclusive measurements of these observables have a huge advantage over those performed with exclusive decays, which is potential signal yields: The inclusive branching fraction is a factor $\sim$10 higher than the $\Bp\to\Kp\mumu$ exclusive branching fraction~\cite{Bailey:2015dka,Du:2015tda}. This in addition to the fact that all $b$-hadron species will produce a \bKll signature, which increase the signal yields by another factor of $\sim$\,2.5. After accounting for an approximate 50\% loss due to the selection of a charged kaon from isospin assumptions, the improvement in signal yields is approximately a factor 15. For example, one should expect 1M \bKll candidates with the dataset expected to be collected at the LHCb experiment by the end of Run III~\cite{Aaij:2636441}. These yields are approximated assuming that the yields in Refs~\cite{Aaij:2014pli,Aaij:2019nmj,Aaij:2021vac} scale with the expected luminosity of 23\invfb.  

The large signal yields mean that there is a lot of room in sensitivity in which to remove backgrounds, particularly if hadronic uncertainties are not an issue. For example, one can completely veto the double-semileptonic background by requiring that $m_{K^{+}\ell^{-}} > m_{D}$. Provided that this is done for both electrons and muons, one avoids the theoretical extrapolations needed to interpret the result. Another tight selection which can be applied is that on the mass of the strange meson, $m_{X_{s}}$. A tight selection here would be detrimental to the systematic uncertainty for the branching fraction but for lepton symmetry and $\CP$ measurements, there is no danger in removing a large fraction of the signal to have a clean sample. Even with a very low signal efficiency of 10\%, this would still represent the biggest signal yields and would potentially provide the most precise measurements for these observables. 

\section{Conclusion}
In summary, a new method to reconstruct inclusive \bsll decays is presented. The method relies on isopsin symmetry to extrapolate the decay \bKll to the fully inclusive rate in \Bp and \Bz decays. This method has the potential to provide measurements with complementary experimental uncertainties with respect to existing ones. 

The fast simulation performed confirms a few experimental advantages expected with respect to a fully inclusive approach. Given that the interpretation of inclusive \incl decays is limited by the experimental uncertainties, any improvement to the experimental signature could prove valuable in complementing the existing exclusive measurements.

At a hadron collider, additional complications arise from the final state diversity of \Bs and \Lb decays. For the branching fraction, we propose to treat these decays as background and subtract them from the decay rate of \Bz and \Bp mesons which forms the signal. Such a treatment will undoubtedly require theoretical input but will be orthogonal to those assumed in a sum-of-exclusives approach. It should be noted that at Belle-II such complications are not present, and therefore such an isospin extrapolation should be well suited if similar experimental advantages are seen in at an \epem collider. 

Regardless of issues related to the extrapolation, an inclusive \bKll signature would provide the world's largest supply of self-tagged \bsll decays. If the background can be controlled, such a sample could provide world leading measurements of theoretically precise observables such as \CP and lepton symmetries. Given that most measurements are statistically limited, the potential increase in signal yield could have a significant impact on the experimental landscape in \bsll decays.

\begin{acknowledgements}
We would like to thank Ulrik Egede for several fruitful discussions when this idea first arose. 
\end{acknowledgements}

\bibliographystyle{LHCb}
\bibliography{references}

\ifx\mcitethebibliography\mciteundefinedmacro
\PackageError{LHCb.bst}{mciteplus.sty has not been loaded}
{This bibstyle requires the use of the mciteplus package.}\fi
\providecommand{\href}[2]{#2}
\begin{mcitethebibliography}{10}
\mciteSetBstSublistMode{n}
\mciteSetBstMaxWidthForm{subitem}{\alph{mcitesubitemcount})}
\mciteSetBstSublistLabelBeginEnd{\mcitemaxwidthsubitemform\space}
{\relax}{\relax}

\bibitem{Aaij:2013qta}
LHCb, R.~Aaij {\em et~al.},
  \ifthenelse{\boolean{articletitles}}{\emph{{Measurement of
  Form-Factor-Independent Observables in the Decay $B^{0} \to K^{*0} \mu^+
  \mu^-$}}, }{}\href{https://doi.org/10.1103/PhysRevLett.111.191801}{Phys.\
  Rev.\ Lett.\  \textbf{111} (2013) 191801},
  \href{http://arxiv.org/abs/1308.1707}{{\normalfont\ttfamily
  arXiv:1308.1707}}\relax
\mciteBstWouldAddEndPuncttrue
\mciteSetBstMidEndSepPunct{\mcitedefaultmidpunct}
{\mcitedefaultendpunct}{\mcitedefaultseppunct}\relax
\EndOfBibitem
\bibitem{Aaij:2014ora}
LHCb, R.~Aaij {\em et~al.}, \ifthenelse{\boolean{articletitles}}{\emph{{Test of
  lepton universality using $B^{+}\rightarrow K^{+}\ell^{+}\ell^{-}$ decays}},
  }{}\href{https://doi.org/10.1103/PhysRevLett.113.151601}{Phys.\ Rev.\ Lett.\
  \textbf{113} (2014) 151601},
  \href{http://arxiv.org/abs/1406.6482}{{\normalfont\ttfamily
  arXiv:1406.6482}}\relax
\mciteBstWouldAddEndPuncttrue
\mciteSetBstMidEndSepPunct{\mcitedefaultmidpunct}
{\mcitedefaultendpunct}{\mcitedefaultseppunct}\relax
\EndOfBibitem
\bibitem{CMS:2014xfa}
CMS, LHCb, V.~Khachatryan {\em et~al.},
  \ifthenelse{\boolean{articletitles}}{\emph{{Observation of the rare
  $B^0_s\to\mu^+\mu^-$ decay from the combined analysis of CMS and LHCb data}},
  }{}\href{https://doi.org/10.1038/nature14474}{Nature \textbf{522} (2015) 68},
  \href{http://arxiv.org/abs/1411.4413}{{\normalfont\ttfamily
  arXiv:1411.4413}}\relax
\mciteBstWouldAddEndPuncttrue
\mciteSetBstMidEndSepPunct{\mcitedefaultmidpunct}
{\mcitedefaultendpunct}{\mcitedefaultseppunct}\relax
\EndOfBibitem
\bibitem{Aaij:2015oid}
LHCb, R.~Aaij {\em et~al.}, \ifthenelse{\boolean{articletitles}}{\emph{{Angular
  analysis of the $B^{0} \to K^{*0} \mu^{+} \mu^{-}$ decay using 3 fb$^{-1}$ of
  integrated luminosity}},
  }{}\href{https://doi.org/10.1007/JHEP02(2016)104}{JHEP \textbf{02} (2016)
  104}, \href{http://arxiv.org/abs/1512.04442}{{\normalfont\ttfamily
  arXiv:1512.04442}}\relax
\mciteBstWouldAddEndPuncttrue
\mciteSetBstMidEndSepPunct{\mcitedefaultmidpunct}
{\mcitedefaultendpunct}{\mcitedefaultseppunct}\relax
\EndOfBibitem
\bibitem{Aaij:2017vbb}
LHCb, R.~Aaij {\em et~al.}, \ifthenelse{\boolean{articletitles}}{\emph{{Test of
  lepton universality with $B^{0} \rightarrow K^{*0}\ell^{+}\ell^{-}$ decays}},
  }{}\href{https://doi.org/10.1007/JHEP08(2017)055}{JHEP \textbf{08} (2017)
  055}, \href{http://arxiv.org/abs/1705.05802}{{\normalfont\ttfamily
  arXiv:1705.05802}}\relax
\mciteBstWouldAddEndPuncttrue
\mciteSetBstMidEndSepPunct{\mcitedefaultmidpunct}
{\mcitedefaultendpunct}{\mcitedefaultseppunct}\relax
\EndOfBibitem
\bibitem{Aaboud:2018mst}
ATLAS, M.~Aaboud {\em et~al.},
  \ifthenelse{\boolean{articletitles}}{\emph{{Study of the rare decays of
  $B^0_s$ and $B^0$ mesons into muon pairs using data collected during 2015 and
  2016 with the ATLAS detector}},
  }{}\href{https://doi.org/10.1007/JHEP04(2019)098}{JHEP \textbf{04} (2019)
  098}, \href{http://arxiv.org/abs/1812.03017}{{\normalfont\ttfamily
  arXiv:1812.03017}}\relax
\mciteBstWouldAddEndPuncttrue
\mciteSetBstMidEndSepPunct{\mcitedefaultmidpunct}
{\mcitedefaultendpunct}{\mcitedefaultseppunct}\relax
\EndOfBibitem
\bibitem{Sirunyan:2019xdu}
CMS, A.~M. Sirunyan {\em et~al.},
  \ifthenelse{\boolean{articletitles}}{\emph{{Measurement of properties of
  B$^0_\mathrm{s}\to\mu^+\mu^-$ decays and search for B$^0\to\mu^+\mu^-$ with
  the CMS experiment}}, }{}\href{https://doi.org/10.1007/JHEP04(2020)188}{JHEP
  \textbf{04} (2020) 188},
  \href{http://arxiv.org/abs/1910.12127}{{\normalfont\ttfamily
  arXiv:1910.12127}}\relax
\mciteBstWouldAddEndPuncttrue
\mciteSetBstMidEndSepPunct{\mcitedefaultmidpunct}
{\mcitedefaultendpunct}{\mcitedefaultseppunct}\relax
\EndOfBibitem
\bibitem{Aaij:2019wad}
LHCb, R.~Aaij {\em et~al.}, \ifthenelse{\boolean{articletitles}}{\emph{{Search
  for lepton-universality violation in $B^+\to K^+\ell^+\ell^-$ decays}},
  }{}\href{https://doi.org/10.1103/PhysRevLett.122.191801}{Phys.\ Rev.\ Lett.\
  \textbf{122} (2019) 191801},
  \href{http://arxiv.org/abs/1903.09252}{{\normalfont\ttfamily
  arXiv:1903.09252}}\relax
\mciteBstWouldAddEndPuncttrue
\mciteSetBstMidEndSepPunct{\mcitedefaultmidpunct}
{\mcitedefaultendpunct}{\mcitedefaultseppunct}\relax
\EndOfBibitem
\bibitem{Aaij:2020nrf}
LHCb, R.~Aaij {\em et~al.},
  \ifthenelse{\boolean{articletitles}}{\emph{{Measurement of $CP$-Averaged
  Observables in the $B^{0}\rightarrow K^{*0}\mu^{+}\mu^{-}$ Decay}},
  }{}\href{https://doi.org/10.1103/PhysRevLett.125.011802}{Phys.\ Rev.\ Lett.\
  \textbf{125} (2020) 011802},
  \href{http://arxiv.org/abs/2003.04831}{{\normalfont\ttfamily
  arXiv:2003.04831}}\relax
\mciteBstWouldAddEndPuncttrue
\mciteSetBstMidEndSepPunct{\mcitedefaultmidpunct}
{\mcitedefaultendpunct}{\mcitedefaultseppunct}\relax
\EndOfBibitem
\bibitem{Aaij:2020ruw}
LHCb, R.~Aaij {\em et~al.}, \ifthenelse{\boolean{articletitles}}{\emph{{Angular
  analysis of the $B^{+}\rightarrow K^{\ast+}\mu^{+}\mu^{-}$ decay}},
  }{}\href{http://arxiv.org/abs/2012.13241}{{\normalfont\ttfamily
  arXiv:2012.13241}}\relax
\mciteBstWouldAddEndPuncttrue
\mciteSetBstMidEndSepPunct{\mcitedefaultmidpunct}
{\mcitedefaultendpunct}{\mcitedefaultseppunct}\relax
\EndOfBibitem
\bibitem{Aaij:2021vac}
LHCb, R.~Aaij {\em et~al.}, \ifthenelse{\boolean{articletitles}}{\emph{{Test of
  lepton universality in beauty-quark decays}},
  }{}\href{http://arxiv.org/abs/2103.11769}{{\normalfont\ttfamily
  arXiv:2103.11769}}\relax
\mciteBstWouldAddEndPuncttrue
\mciteSetBstMidEndSepPunct{\mcitedefaultmidpunct}
{\mcitedefaultendpunct}{\mcitedefaultseppunct}\relax
\EndOfBibitem
\bibitem{Alguero:2021anc}
M.~Alguer\'o {\em et~al.},
  \ifthenelse{\boolean{articletitles}}{\emph{$\boldsymbol{b\to s\ell\ell}$
  global fits after moriond 2021 results},
  }{}\href{http://arxiv.org/abs/2104.08921}{{\normalfont\ttfamily
  arXiv:2104.08921}}\relax
\mciteBstWouldAddEndPuncttrue
\mciteSetBstMidEndSepPunct{\mcitedefaultmidpunct}
{\mcitedefaultendpunct}{\mcitedefaultseppunct}\relax
\EndOfBibitem
\bibitem{Altmannshofer:2021qrr}
W.~Altmannshofer and P.~Stangl, \ifthenelse{\boolean{articletitles}}{\emph{{New
  Physics in Rare B Decays after Moriond 2021}},
  }{}\href{http://arxiv.org/abs/2103.13370}{{\normalfont\ttfamily
  arXiv:2103.13370}}\relax
\mciteBstWouldAddEndPuncttrue
\mciteSetBstMidEndSepPunct{\mcitedefaultmidpunct}
{\mcitedefaultendpunct}{\mcitedefaultseppunct}\relax
\EndOfBibitem
\bibitem{Geng:2021nhg}
L.-S. Geng {\em et~al.},
  \ifthenelse{\boolean{articletitles}}{\emph{{Implications of new evidence for
  lepton-universality violation in $b\to s\ell^+\ell^-$ decays}},
  }{}\href{http://arxiv.org/abs/2103.12738}{{\normalfont\ttfamily
  arXiv:2103.12738}}\relax
\mciteBstWouldAddEndPuncttrue
\mciteSetBstMidEndSepPunct{\mcitedefaultmidpunct}
{\mcitedefaultendpunct}{\mcitedefaultseppunct}\relax
\EndOfBibitem
\bibitem{Lyon:2014hpa}
J.~Lyon and R.~Zwicky, \ifthenelse{\boolean{articletitles}}{\emph{{Resonances
  gone topsy turvy - the charm of QCD or new physics in $b \to s \ell^+
  \ell^-$?}}, }{}\href{http://arxiv.org/abs/1406.0566}{{\normalfont\ttfamily
  arXiv:1406.0566}}\relax
\mciteBstWouldAddEndPuncttrue
\mciteSetBstMidEndSepPunct{\mcitedefaultmidpunct}
{\mcitedefaultendpunct}{\mcitedefaultseppunct}\relax
\EndOfBibitem
\bibitem{Bharucha:2020eup}
A.~Bharucha, D.~Boito, and C.~M\'eaux,
  \ifthenelse{\boolean{articletitles}}{\emph{{Disentangling QCD and New Physics
  in $D^+\to\pi^+\ell^+\ell^-$}},
  }{}\href{http://arxiv.org/abs/2011.12856}{{\normalfont\ttfamily
  arXiv:2011.12856}}\relax
\mciteBstWouldAddEndPuncttrue
\mciteSetBstMidEndSepPunct{\mcitedefaultmidpunct}
{\mcitedefaultendpunct}{\mcitedefaultseppunct}\relax
\EndOfBibitem
\bibitem{Gubernari:2020eft}
N.~Gubernari, D.~Van~Dyk, and J.~Virto,
  \ifthenelse{\boolean{articletitles}}{\emph{{Non-local matrix elements in
  $B_{(s)}\to \{K^{(*)},\phi\}\ell^+\ell^-$}},
  }{}\href{https://doi.org/10.1007/JHEP02(2021)088}{JHEP \textbf{02} (2021)
  088}, \href{http://arxiv.org/abs/2011.09813}{{\normalfont\ttfamily
  arXiv:2011.09813}}\relax
\mciteBstWouldAddEndPuncttrue
\mciteSetBstMidEndSepPunct{\mcitedefaultmidpunct}
{\mcitedefaultendpunct}{\mcitedefaultseppunct}\relax
\EndOfBibitem
\bibitem{Huber:2019iqf}
T.~Huber {\em et~al.}, \ifthenelse{\boolean{articletitles}}{\emph{{Long
  distance effects in inclusive rare B decays and phenomenology of $\bar{B}\to
  X_d \ell^+\ell^-$}}, }{}\href{https://doi.org/10.1007/JHEP10(2019)228}{JHEP
  \textbf{10} (2019) 228},
  \href{http://arxiv.org/abs/1908.07507}{{\normalfont\ttfamily
  arXiv:1908.07507}}\relax
\mciteBstWouldAddEndPuncttrue
\mciteSetBstMidEndSepPunct{\mcitedefaultmidpunct}
{\mcitedefaultendpunct}{\mcitedefaultseppunct}\relax
\EndOfBibitem
\bibitem{Jager:2019bgk}
S.~J\"ager, M.~Kirk, A.~Lenz, and K.~Leslie,
  \ifthenelse{\boolean{articletitles}}{\emph{{Charming New $B$-Physics}},
  }{}\href{https://doi.org/10.1007/JHEP03(2020)122}{JHEP \textbf{03} (2020)
  122}, \href{http://arxiv.org/abs/1910.12924}{{\normalfont\ttfamily
  arXiv:1910.12924}}\relax
\mciteBstWouldAddEndPuncttrue
\mciteSetBstMidEndSepPunct{\mcitedefaultmidpunct}
{\mcitedefaultendpunct}{\mcitedefaultseppunct}\relax
\EndOfBibitem
\bibitem{Nakayama:2019eth}
JLQCD, K.~Nakayama and S.~Hashimoto,
  \ifthenelse{\boolean{articletitles}}{\emph{{Test of factorization for the
  long-distance effects from charmonium on $B\to K\ell^+\ell^-$}},
  }{}\href{https://doi.org/10.22323/1.334.0221}{PoS \textbf{LATTICE2018} (2019)
  221}, \href{http://arxiv.org/abs/1901.08784}{{\normalfont\ttfamily
  arXiv:1901.08784}}\relax
\mciteBstWouldAddEndPuncttrue
\mciteSetBstMidEndSepPunct{\mcitedefaultmidpunct}
{\mcitedefaultendpunct}{\mcitedefaultseppunct}\relax
\EndOfBibitem
\bibitem{Arbey:2018ics}
A.~Arbey, T.~Hurth, F.~Mahmoudi, and S.~Neshatpour,
  \ifthenelse{\boolean{articletitles}}{\emph{{Hadronic and New Physics
  Contributions to $b \to s$ Transitions}},
  }{}\href{https://doi.org/10.1103/PhysRevD.98.095027}{Phys.\ Rev.\ D
  \textbf{98} (2018) 095027},
  \href{http://arxiv.org/abs/1806.02791}{{\normalfont\ttfamily
  arXiv:1806.02791}}\relax
\mciteBstWouldAddEndPuncttrue
\mciteSetBstMidEndSepPunct{\mcitedefaultmidpunct}
{\mcitedefaultendpunct}{\mcitedefaultseppunct}\relax
\EndOfBibitem
\bibitem{Chrzaszcz:2018yza}
M.~Chrzaszcz {\em et~al.},
  \ifthenelse{\boolean{articletitles}}{\emph{{Prospects for disentangling long-
  and short-distance effects in the decays $B\to K^* \mu^+\mu^-$}},
  }{}\href{https://doi.org/10.1007/JHEP10(2019)236}{JHEP \textbf{10} (2019)
  236}, \href{http://arxiv.org/abs/1805.06378}{{\normalfont\ttfamily
  arXiv:1805.06378}}\relax
\mciteBstWouldAddEndPuncttrue
\mciteSetBstMidEndSepPunct{\mcitedefaultmidpunct}
{\mcitedefaultendpunct}{\mcitedefaultseppunct}\relax
\EndOfBibitem
\bibitem{Blake:2017fyh}
T.~Blake {\em et~al.}, \ifthenelse{\boolean{articletitles}}{\emph{{An empirical
  model to determine the hadronic resonance contributions to $\overline{B}{} ^0
  \!\rightarrow \overline{K}{} ^{*0} \mu ^+ \mu ^- $ transitions}},
  }{}\href{https://doi.org/10.1140/epjc/s10052-018-5937-3}{Eur.\ Phys.\ J.\ C
  \textbf{78} (2018) 453},
  \href{http://arxiv.org/abs/1709.03921}{{\normalfont\ttfamily
  arXiv:1709.03921}}\relax
\mciteBstWouldAddEndPuncttrue
\mciteSetBstMidEndSepPunct{\mcitedefaultmidpunct}
{\mcitedefaultendpunct}{\mcitedefaultseppunct}\relax
\EndOfBibitem
\bibitem{Ahmady:2015fha}
M.~Ahmady, D.~Hatfield, S.~Lord, and R.~Sandapen,
  \ifthenelse{\boolean{articletitles}}{\emph{{Effect of $c\overline{c}$
  resonances in the branching ratio and forward-backward asymmetry of the decay
  $B \to K^{*} \mu^+ \mu^-$}},
  }{}\href{https://doi.org/10.1103/PhysRevD.92.114028}{Phys.\ Rev.\ D
  \textbf{92} (2015) 114028},
  \href{http://arxiv.org/abs/1508.02327}{{\normalfont\ttfamily
  arXiv:1508.02327}}\relax
\mciteBstWouldAddEndPuncttrue
\mciteSetBstMidEndSepPunct{\mcitedefaultmidpunct}
{\mcitedefaultendpunct}{\mcitedefaultseppunct}\relax
\EndOfBibitem
\bibitem{Descotes-Genon:2015xqa}
S.~Descotes-Genon, L.~Hofer, J.~Matias, and J.~Virto,
  \ifthenelse{\boolean{articletitles}}{\emph{{Theoretical status of $B \to K^*
  \mu^+\mu^-$: The path towards New Physics}},
  }{}\href{https://doi.org/10.1088/1742-6596/631/1/012027}{J.\ Phys.\ Conf.\
  Ser.\  \textbf{631} (2015) 012027},
  \href{http://arxiv.org/abs/1503.03328}{{\normalfont\ttfamily
  arXiv:1503.03328}}\relax
\mciteBstWouldAddEndPuncttrue
\mciteSetBstMidEndSepPunct{\mcitedefaultmidpunct}
{\mcitedefaultendpunct}{\mcitedefaultseppunct}\relax
\EndOfBibitem
\bibitem{Ciuchini:2015qxb}
M.~Ciuchini {\em et~al.}, \ifthenelse{\boolean{articletitles}}{\emph{{$B\to K^*
  \ell^+ \ell^-$ decays at large recoil in the Standard Model: a theoretical
  reappraisal}}, }{}\href{https://doi.org/10.1007/JHEP06(2016)116}{JHEP
  \textbf{06} (2016) 116},
  \href{http://arxiv.org/abs/1512.07157}{{\normalfont\ttfamily
  arXiv:1512.07157}}\relax
\mciteBstWouldAddEndPuncttrue
\mciteSetBstMidEndSepPunct{\mcitedefaultmidpunct}
{\mcitedefaultendpunct}{\mcitedefaultseppunct}\relax
\EndOfBibitem
\bibitem{Aaij:2019bzx}
LHCb, R.~Aaij {\em et~al.}, \ifthenelse{\boolean{articletitles}}{\emph{{Test of
  lepton universality with $ {\Lambda}_b^0\to
  {pK}^{-}{\mathrm{\ell}}^{+}{\mathrm{\ell}}^{-} $ decays}},
  }{}\href{https://doi.org/10.1007/JHEP05(2020)040}{JHEP \textbf{05} (2020)
  040}, \href{http://arxiv.org/abs/1912.08139}{{\normalfont\ttfamily
  arXiv:1912.08139}}\relax
\mciteBstWouldAddEndPuncttrue
\mciteSetBstMidEndSepPunct{\mcitedefaultmidpunct}
{\mcitedefaultendpunct}{\mcitedefaultseppunct}\relax
\EndOfBibitem
\bibitem{Aaij:2017vad}
LHCb, R.~Aaij {\em et~al.},
  \ifthenelse{\boolean{articletitles}}{\emph{{Measurement of the
  $B^0_s\to\mu^+\mu^-$ branching fraction and effective lifetime and search for
  $B^0\to\mu^+\mu^-$ decays}},
  }{}\href{https://doi.org/10.1103/PhysRevLett.118.191801}{Phys.\ Rev.\ Lett.\
  \textbf{118} (2017) 191801},
  \href{http://arxiv.org/abs/1703.05747}{{\normalfont\ttfamily
  arXiv:1703.05747}}\relax
\mciteBstWouldAddEndPuncttrue
\mciteSetBstMidEndSepPunct{\mcitedefaultmidpunct}
{\mcitedefaultendpunct}{\mcitedefaultseppunct}\relax
\EndOfBibitem
\bibitem{Lancierini:2021sdf}
D.~Lancierini, G.~Isidori, P.~Owen, and N.~Serra,
  \ifthenelse{\boolean{articletitles}}{\emph{{On the significance of new
  physics in $b\to s\ell^+\ell^-$ decays}},
  }{}\href{http://arxiv.org/abs/2104.05631}{{\normalfont\ttfamily
  arXiv:2104.05631}}\relax
\mciteBstWouldAddEndPuncttrue
\mciteSetBstMidEndSepPunct{\mcitedefaultmidpunct}
{\mcitedefaultendpunct}{\mcitedefaultseppunct}\relax
\EndOfBibitem
\bibitem{Hurth:2021nsi}
T.~Hurth, F.~Mahmoudi, D.~M. Santos, and S.~Neshatpour,
  \ifthenelse{\boolean{articletitles}}{\emph{{More Indications for Lepton
  Nonuniversality in $b \to s \ell^+ \ell^-$}},
  }{}\href{http://arxiv.org/abs/2104.10058}{{\normalfont\ttfamily
  arXiv:2104.10058}}\relax
\mciteBstWouldAddEndPuncttrue
\mciteSetBstMidEndSepPunct{\mcitedefaultmidpunct}
{\mcitedefaultendpunct}{\mcitedefaultseppunct}\relax
\EndOfBibitem
\bibitem{Cornella:2021sby}
C.~Cornella {\em et~al.}, \ifthenelse{\boolean{articletitles}}{\emph{{Reading
  the footprints of the B-meson flavor anomalies}},
  }{}\href{http://arxiv.org/abs/2103.16558}{{\normalfont\ttfamily
  arXiv:2103.16558}}\relax
\mciteBstWouldAddEndPuncttrue
\mciteSetBstMidEndSepPunct{\mcitedefaultmidpunct}
{\mcitedefaultendpunct}{\mcitedefaultseppunct}\relax
\EndOfBibitem
\bibitem{Straub:2015ica}
A.~Bharucha, D.~M. Straub, and R.~Zwicky,
  \ifthenelse{\boolean{articletitles}}{\emph{{$B\to V\ell^+\ell^-$ in the
  Standard Model from light-cone sum rules}},
  }{}\href{https://doi.org/10.1007/JHEP08(2016)098}{JHEP \textbf{08} (2016)
  098}, \href{http://arxiv.org/abs/1503.05534}{{\normalfont\ttfamily
  arXiv:1503.05534}}\relax
\mciteBstWouldAddEndPuncttrue
\mciteSetBstMidEndSepPunct{\mcitedefaultmidpunct}
{\mcitedefaultendpunct}{\mcitedefaultseppunct}\relax
\EndOfBibitem
\bibitem{Bailey:2015dka}
J.~A. Bailey {\em et~al.}, \ifthenelse{\boolean{articletitles}}{\emph{{$B\to
  Kl^+l^-$ decay form factors from three-flavor lattice QCD}},
  }{}\href{https://doi.org/10.1103/PhysRevD.93.025026}{Phys.\ Rev.\
  \textbf{D93} (2016) 025026},
  \href{http://arxiv.org/abs/1509.06235}{{\normalfont\ttfamily
  arXiv:1509.06235}}\relax
\mciteBstWouldAddEndPuncttrue
\mciteSetBstMidEndSepPunct{\mcitedefaultmidpunct}
{\mcitedefaultendpunct}{\mcitedefaultseppunct}\relax
\EndOfBibitem
\bibitem{Du:2015tda}
D.~Du {\em et~al.}, \ifthenelse{\boolean{articletitles}}{\emph{{Phenomenology
  of semileptonic B-meson decays with form factors from lattice QCD}},
  }{}\href{https://doi.org/10.1103/PhysRevD.93.034005}{Phys.\ Rev.\
  \textbf{D93} (2016) 034005},
  \href{http://arxiv.org/abs/1510.02349}{{\normalfont\ttfamily
  arXiv:1510.02349}}\relax
\mciteBstWouldAddEndPuncttrue
\mciteSetBstMidEndSepPunct{\mcitedefaultmidpunct}
{\mcitedefaultendpunct}{\mcitedefaultseppunct}\relax
\EndOfBibitem
\bibitem{Horgan:2013pva}
R.~R. Horgan, Z.~Liu, S.~Meinel, and M.~Wingate,
  \ifthenelse{\boolean{articletitles}}{\emph{{Calculation of $B^0 \to K^{*0}
  \mu^+ \mu^-$ and $B_s^0 \to \phi \mu^+ \mu^-$ observables using form factors
  from lattice QCD}},
  }{}\href{https://doi.org/10.1103/PhysRevLett.112.212003}{Phys.\ Rev.\ Lett.\
  \textbf{112} (2014) 212003},
  \href{http://arxiv.org/abs/1310.3887}{{\normalfont\ttfamily
  arXiv:1310.3887}}\relax
\mciteBstWouldAddEndPuncttrue
\mciteSetBstMidEndSepPunct{\mcitedefaultmidpunct}
{\mcitedefaultendpunct}{\mcitedefaultseppunct}\relax
\EndOfBibitem
\bibitem{Horgan:2015vla}
R.~R. Horgan, Z.~Liu, S.~Meinel, and M.~Wingate,
  \ifthenelse{\boolean{articletitles}}{\emph{{Rare $B$ decays using lattice QCD
  form factors}}, }{}\href{https://doi.org/10.22323/1.214.0372}{PoS
  \textbf{LATTICE2014} (2015) 372},
  \href{http://arxiv.org/abs/1501.00367}{{\normalfont\ttfamily
  arXiv:1501.00367}}\relax
\mciteBstWouldAddEndPuncttrue
\mciteSetBstMidEndSepPunct{\mcitedefaultmidpunct}
{\mcitedefaultendpunct}{\mcitedefaultseppunct}\relax
\EndOfBibitem
\bibitem{Aaij:2014pli}
LHCb, R.~Aaij {\em et~al.},
  \ifthenelse{\boolean{articletitles}}{\emph{{Differential branching fractions
  and isospin asymmetries of $B \to K^{(*)} \mu^+ \mu^-$ decays}},
  }{}\href{https://doi.org/10.1007/JHEP06(2014)133}{JHEP \textbf{06} (2014)
  133}, \href{http://arxiv.org/abs/1403.8044}{{\normalfont\ttfamily
  arXiv:1403.8044}}\relax
\mciteBstWouldAddEndPuncttrue
\mciteSetBstMidEndSepPunct{\mcitedefaultmidpunct}
{\mcitedefaultendpunct}{\mcitedefaultseppunct}\relax
\EndOfBibitem
\bibitem{Aaij:2016flj}
LHCb, R.~Aaij {\em et~al.},
  \ifthenelse{\boolean{articletitles}}{\emph{{Measurements of the S-wave
  fraction in $B^{0}\rightarrow K^{+}\pi^{-}\mu^{+}\mu^{-}$ decays and the
  $B^{0}\rightarrow K^{\ast}(892)^{0}\mu^{+}\mu^{-}$ differential branching
  fraction}}, }{}\href{https://doi.org/10.1007/JHEP11(2016)047}{JHEP
  \textbf{11} (2016) 047},
  \href{http://arxiv.org/abs/1606.04731}{{\normalfont\ttfamily
  arXiv:1606.04731}}, [Erratum: JHEP 04, 142 (2017)]\relax
\mciteBstWouldAddEndPuncttrue
\mciteSetBstMidEndSepPunct{\mcitedefaultmidpunct}
{\mcitedefaultendpunct}{\mcitedefaultseppunct}\relax
\EndOfBibitem
\bibitem{Aaij:2021pkz}
LHCb, R.~Aaij {\em et~al.},
  \ifthenelse{\boolean{articletitles}}{\emph{{Branching fraction measurements
  of the rare $B^0_s\rightarrow\phi\mu^+\mu^-$ and $B^0_s\rightarrow
  f_2^\prime(1525)\mu^+\mu^-$ decays}},
  }{}\href{http://arxiv.org/abs/2105.14007}{{\normalfont\ttfamily
  arXiv:2105.14007}}\relax
\mciteBstWouldAddEndPuncttrue
\mciteSetBstMidEndSepPunct{\mcitedefaultmidpunct}
{\mcitedefaultendpunct}{\mcitedefaultseppunct}\relax
\EndOfBibitem
\bibitem{Hurth:2010tk}
T.~Hurth and M.~Nakao, \ifthenelse{\boolean{articletitles}}{\emph{{Radiative
  and Electroweak Penguin Decays of B Mesons}},
  }{}\href{https://doi.org/10.1146/annurev.nucl.012809.104424}{Ann.\ Rev.\
  Nucl.\ Part.\ Sci.\  \textbf{60} (2010) 645},
  \href{http://arxiv.org/abs/1005.1224}{{\normalfont\ttfamily
  arXiv:1005.1224}}\relax
\mciteBstWouldAddEndPuncttrue
\mciteSetBstMidEndSepPunct{\mcitedefaultmidpunct}
{\mcitedefaultendpunct}{\mcitedefaultseppunct}\relax
\EndOfBibitem
\bibitem{Huber:2015sra}
T.~Huber, T.~Hurth, and E.~Lunghi,
  \ifthenelse{\boolean{articletitles}}{\emph{{Inclusive $ \overline{B}\to
  {X}_s{\ell}^{+}{\ell}^{-} $ : complete angular analysis and a thorough study
  of collinear photons}},
  }{}\href{https://doi.org/10.1007/JHEP06(2015)176}{JHEP \textbf{06} (2015)
  176}, \href{http://arxiv.org/abs/1503.04849}{{\normalfont\ttfamily
  arXiv:1503.04849}}\relax
\mciteBstWouldAddEndPuncttrue
\mciteSetBstMidEndSepPunct{\mcitedefaultmidpunct}
{\mcitedefaultendpunct}{\mcitedefaultseppunct}\relax
\EndOfBibitem
\bibitem{Huber:2020vup}
T.~Huber {\em et~al.},
  \ifthenelse{\boolean{articletitles}}{\emph{{Phenomenology of inclusive $
  \overline{B}\to {X}_s{\mathrm{\ell}}^{+}{\mathrm{\ell}}^{-} $ for the Belle
  II era}}, }{}\href{https://doi.org/10.1007/JHEP10(2020)088}{JHEP \textbf{10}
  (2020) 088}, \href{http://arxiv.org/abs/2007.04191}{{\normalfont\ttfamily
  arXiv:2007.04191}}\relax
\mciteBstWouldAddEndPuncttrue
\mciteSetBstMidEndSepPunct{\mcitedefaultmidpunct}
{\mcitedefaultendpunct}{\mcitedefaultseppunct}\relax
\EndOfBibitem
\bibitem{Lees:2013nxa}
BaBar, J.~P. Lees {\em et~al.},
  \ifthenelse{\boolean{articletitles}}{\emph{{Measurement of the $B \to X_s
  l^+l^-$ branching fraction and search for direct CP violation from a sum of
  exclusive final states}},
  }{}\href{https://doi.org/10.1103/PhysRevLett.112.211802}{Phys.\ Rev.\ Lett.\
  \textbf{112} (2014) 211802},
  \href{http://arxiv.org/abs/1312.5364}{{\normalfont\ttfamily
  arXiv:1312.5364}}\relax
\mciteBstWouldAddEndPuncttrue
\mciteSetBstMidEndSepPunct{\mcitedefaultmidpunct}
{\mcitedefaultendpunct}{\mcitedefaultseppunct}\relax
\EndOfBibitem
\bibitem{Iwasaki:2005sy}
Belle, M.~Iwasaki {\em et~al.},
  \ifthenelse{\boolean{articletitles}}{\emph{{Improved measurement of the
  electroweak penguin process $B \to X_s l^+ l^-$}},
  }{}\href{https://doi.org/10.1103/PhysRevD.72.092005}{Phys.\ Rev.\ D
  \textbf{72} (2005) 092005},
  \href{http://arxiv.org/abs/hep-ex/0503044}{{\normalfont\ttfamily
  arXiv:hep-ex/0503044}}\relax
\mciteBstWouldAddEndPuncttrue
\mciteSetBstMidEndSepPunct{\mcitedefaultmidpunct}
{\mcitedefaultendpunct}{\mcitedefaultseppunct}\relax
\EndOfBibitem
\bibitem{Sato:2014pjr}
Belle, Y.~Sato {\em et~al.},
  \ifthenelse{\boolean{articletitles}}{\emph{{Measurement of the lepton
  forward-backward asymmetry in $B \rightarrow X_s \ell^+ \ell^-$ decays with a
  sum of exclusive modes}},
  }{}\href{https://doi.org/10.1103/PhysRevD.93.059901}{Phys.\ Rev.\ D
  \textbf{93} (2016) 032008},
  \href{http://arxiv.org/abs/1402.7134}{{\normalfont\ttfamily
  arXiv:1402.7134}}, [Addendum: Phys.Rev.D 93, 059901 (2016)]\relax
\mciteBstWouldAddEndPuncttrue
\mciteSetBstMidEndSepPunct{\mcitedefaultmidpunct}
{\mcitedefaultendpunct}{\mcitedefaultseppunct}\relax
\EndOfBibitem
\bibitem{Kou:2018nap}
Belle-II, W.~Altmannshofer {\em et~al.},
  \ifthenelse{\boolean{articletitles}}{\emph{{The Belle II Physics Book}},
  }{}\href{https://doi.org/10.1093/ptep/ptz106}{PTEP \textbf{2019} (2019)
  123C01}, \href{http://arxiv.org/abs/1808.10567}{{\normalfont\ttfamily
  arXiv:1808.10567}}, [Erratum: PTEP 2020, 029201 (2020)]\relax
\mciteBstWouldAddEndPuncttrue
\mciteSetBstMidEndSepPunct{\mcitedefaultmidpunct}
{\mcitedefaultendpunct}{\mcitedefaultseppunct}\relax
\EndOfBibitem
\bibitem{Cowan:2016tnm}
G.~A. Cowan, D.~C. Craik, and M.~D. Needham,
  \ifthenelse{\boolean{articletitles}}{\emph{{RapidSim: an application for the
  fast simulation of heavy-quark hadron decays}},
  }{}\href{https://doi.org/10.1016/j.cpc.2017.01.029}{Comput.\ Phys.\ Commun.\
  \textbf{214} (2017) 239},
  \href{http://arxiv.org/abs/1612.07489}{{\normalfont\ttfamily
  arXiv:1612.07489}}\relax
\mciteBstWouldAddEndPuncttrue
\mciteSetBstMidEndSepPunct{\mcitedefaultmidpunct}
{\mcitedefaultendpunct}{\mcitedefaultseppunct}\relax
\EndOfBibitem
\bibitem{Aaij:2015yra}
LHCb, R.~Aaij {\em et~al.},
  \ifthenelse{\boolean{articletitles}}{\emph{{Measurement of the ratio of
  branching fractions $\mathcal{B}(\bar{B}^0 \to
  D^{*+}\tau^{-}\bar{\nu}_{\tau})/\mathcal{B}(\bar{B}^0 \to
  D^{*+}\mu^{-}\bar{\nu}_{\mu})$}},
  }{}\href{https://doi.org/https://doi.org/10.1103/PhysRevLett.115.111803}{Phys.\
  Rev.\ Lett.\  \textbf{115} (2015) 111803},
  \href{http://arxiv.org/abs/1506.08614}{{\normalfont\ttfamily
  arXiv:1506.08614}}, [Erratum: Phys.Rev.Lett. 115, 159901 (2015)]\relax
\mciteBstWouldAddEndPuncttrue
\mciteSetBstMidEndSepPunct{\mcitedefaultmidpunct}
{\mcitedefaultendpunct}{\mcitedefaultseppunct}\relax
\EndOfBibitem
\bibitem{Khodjamirian:2012rm}
A.~Khodjamirian, T.~Mannel, and Y.~M. Wang,
  \ifthenelse{\boolean{articletitles}}{\emph{{$B \to K \ell^{+}\ell^{-}$ decay
  at large hadronic recoil}},
  }{}\href{https://doi.org/10.1007/JHEP02(2013)010}{JHEP \textbf{02} (2013)
  010}, \href{http://arxiv.org/abs/1211.0234}{{\normalfont\ttfamily
  arXiv:1211.0234}}\relax
\mciteBstWouldAddEndPuncttrue
\mciteSetBstMidEndSepPunct{\mcitedefaultmidpunct}
{\mcitedefaultendpunct}{\mcitedefaultseppunct}\relax
\EndOfBibitem
\bibitem{Lyon:2013gba}
J.~Lyon and R.~Zwicky, \ifthenelse{\boolean{articletitles}}{\emph{{Isospin
  asymmetries in $B\to(K^*,\rho)\gamma/l^+l^-$ and $B\to Kl^+l^-$ in and beyond
  the standard model}},
  }{}\href{https://doi.org/10.1103/PhysRevD.88.094004}{Phys.\ Rev.\ D
  \textbf{88} (2013) 094004},
  \href{http://arxiv.org/abs/1305.4797}{{\normalfont\ttfamily
  arXiv:1305.4797}}\relax
\mciteBstWouldAddEndPuncttrue
\mciteSetBstMidEndSepPunct{\mcitedefaultmidpunct}
{\mcitedefaultendpunct}{\mcitedefaultseppunct}\relax
\EndOfBibitem
\bibitem{Aaij:2019pqz}
LHCb, R.~Aaij {\em et~al.},
  \ifthenelse{\boolean{articletitles}}{\emph{{Measurement of $b$ hadron
  fractions in 13 TeV $pp$ collisions}},
  }{}\href{https://doi.org/10.1103/PhysRevD.100.031102}{Phys.\ Rev.\ D
  \textbf{100} (2019) 031102},
  \href{http://arxiv.org/abs/1902.06794}{{\normalfont\ttfamily
  arXiv:1902.06794}}\relax
\mciteBstWouldAddEndPuncttrue
\mciteSetBstMidEndSepPunct{\mcitedefaultmidpunct}
{\mcitedefaultendpunct}{\mcitedefaultseppunct}\relax
\EndOfBibitem
\bibitem{Aaij:2019vvl}
LHCb, R.~Aaij {\em et~al.},
  \ifthenelse{\boolean{articletitles}}{\emph{{Measurement of the electron
  reconstruction efficiency at LHCb}},
  }{}\href{https://doi.org/10.1088/1748-0221/14/11/P11023}{JINST \textbf{14}
  (2019) P11023}, \href{http://arxiv.org/abs/1909.02957}{{\normalfont\ttfamily
  arXiv:1909.02957}}\relax
\mciteBstWouldAddEndPuncttrue
\mciteSetBstMidEndSepPunct{\mcitedefaultmidpunct}
{\mcitedefaultendpunct}{\mcitedefaultseppunct}\relax
\EndOfBibitem
\bibitem{Aaij:2636441}
LHCb Collaboration, R.~Aaij {\em et~al.},
  \ifthenelse{\boolean{articletitles}}{\emph{{Physics case for an LHCb Upgrade
  II - Opportunities in flavour physics, and beyond, in the HL-LHC era}}, }{}
  tech. rep., CERN, Geneva, 2018.
\newblock ISBN 978-92-9083-494-6,
  doi:~\href{https://doi.org/10347/15157}{10347/15157}\relax
\mciteBstWouldAddEndPuncttrue
\mciteSetBstMidEndSepPunct{\mcitedefaultmidpunct}
{\mcitedefaultendpunct}{\mcitedefaultseppunct}\relax
\EndOfBibitem
\bibitem{Aaij:2019nmj}
LHCb, R.~Aaij {\em et~al.}, \ifthenelse{\boolean{articletitles}}{\emph{{Search
  for Lepton-Flavor Violating Decays $B^+ \to K^+ {\mu}^{\pm} e^{\mp}$}},
  }{}\href{https://doi.org/10.1103/PhysRevLett.123.241802}{Phys.\ Rev.\ Lett.\
  \textbf{123} (2019) 241802},
  \href{http://arxiv.org/abs/1909.01010}{{\normalfont\ttfamily
  arXiv:1909.01010}}\relax
\mciteBstWouldAddEndPuncttrue
\mciteSetBstMidEndSepPunct{\mcitedefaultmidpunct}
{\mcitedefaultendpunct}{\mcitedefaultseppunct}\relax
\EndOfBibitem
\end{mcitethebibliography}

\end{document}